\documentclass[twocolumn,amsmath,amssymb,color,superscriptaddress]{revtex4}
\usepackage{graphicx}
\usepackage{color}

\usepackage{graphicx}% Include figure files
\usepackage{dcolumn}% Align table columns on decimal point
\usepackage{bm}% bold math
\usepackage[T1]{fontenc}
\usepackage{hyperref}
\hypersetup{
     colorlinks = true,
     citecolor  = blue,  % cite
     linkcolor  = magenta % ref
}
\begin{document}

\title{ Dimensional Crossover Tuned by Pressure in Layered Magnetic NiPS$_{3}$}

\author{Xiaoli Ma}
\thanks{These authors contributed equally to this work.}
\affiliation{Beijing National Laboratory for Condensed Matter Physics,
Institute of Physics, Chinese Academy of Sciences, Beijing 100190, China}

\author{Yimeng Wang}
\thanks{These authors contributed equally to this work.}
\affiliation{Department of Physics, Renmin University of China, Beijing 100872, China}

\author{Yunyu Yin}
\thanks{These authors contributed equally to this work.}
\affiliation{Beijing National Laboratory for Condensed Matter Physics,
Institute of Physics, Chinese Academy of Sciences, Beijing 100190, China}

\author{Binbin Yue}
\affiliation{Center for High Pressure Science and Technology Advanced Research, Beijing, 100094, China}

\author{Jianhong Dai}
\affiliation{Beijing National Laboratory for Condensed Matter Physics,
Institute of Physics, Chinese Academy of Sciences, Beijing 100190, China}

\author{Jianting Ji}
\affiliation{Beijing National Laboratory for Condensed Matter Physics,
Institute of Physics, Chinese Academy of Sciences, Beijing 100190, China}

\author{Feng Jin}
\affiliation{Beijing National Laboratory for Condensed Matter Physics,
Institute of Physics, Chinese Academy of Sciences, Beijing 100190, China}
\author{Fang Hong}
\email[e-mail address:]{hongfang@iphy.ac.cn}
\affiliation{Beijing National Laboratory for Condensed Matter Physics,
Institute of Physics, Chinese Academy of Sciences, Beijing 100190, China}

\author{Jian-Tao Wang}
\email[e-mail address:]{ wjt@aphy.iphy.ac.cn}
\affiliation{Beijing National Laboratory for Condensed Matter Physics,
             Institute of Physics, Chinese Academy of Sciences, Beijing 100190, China}
\affiliation{School of Physical Sciences, University of Chinese Academy of Sciences, Beijing 100049, China}
\affiliation{Songshan Lake Materials Laboratory, Dongguan, Guangdong 523808, China}

\author{Qingming Zhang}
\email[e-mail address:]{qmzhang@ruc.edu.cn}
\affiliation{School of Physical Science and Technology, Lanzhou University, Lanzhou 730000, China}
\affiliation{Beijing National Laboratory for Condensed Matter Physics,
Institute of Physics, Chinese Academy of Sciences, Beijing 100190, China}

\author{Xiaohui Yu}
\email[e-mail address:]{yuxh@iphy.ac.cn}
\affiliation{Beijing National Laboratory for Condensed Matter Physics,
Institute of Physics, Chinese Academy of Sciences, Beijing 100190, China}

%\date{\today}

\begin{abstract}
Layered magnetic transition-metal thiophosphate NiPS$_{3}$ has unique two-dimensional (2D) magnetic properties and electronic behavior. 
The electronic band structure and corresponding magnetic state are expected to sensitive to the interlayer interaction, which can be tuned by external pressure. Here, we report an insulator-metal transition accompanied with magnetism collapse during the 2D-3D crossover in structure induced by hydrostatic pressure. A two-stage phase transition from monoclinic ($C2/m$) to trigonal ($P\bar{3}1m$) lattice is identified by $ab$ $initio$ simulation and confirmed by high-pressure XRD and Raman data, corresponding to a layer by layer slip mechanism along the $a$-axis.
Temperature dependence resistance measurements and room temperature infrared spectroscopy show that the insulator-metal transition occurs near 20 GPa as well as magnetism collapse, which is further confirmed by low temperature Raman measurement and theoretical calculation.
These results establish a strong correlation among the structural change, electric transport, and magnetic phase transition and expand our understandings about the layered magnetic materials.

\end{abstract}

\maketitle

\section{INTRODUCTION}
Dimensionality plays an important role to determine the structure, electronic and magnetic properties in most condensed matter systems. The layered materials with van der Waals (vdW) interaction provide a good platform for us to study how the phases and interactions evolve when the material is tuned from the two-dimensional (2D) limit to the there-dimensional (3D) state.
Since graphene was first stripped from the natural graphite by Andre Geim and Konstantin Novoselov \cite{Geim2007},
a large number of 2D materials with atomic thickness then became the new favorite of the scientific community, such as 2D transition metal dichalcogenides \cite{tmd01,tmd02,tmd03}.
Recently, the layered transition metal trichalcogenides (MPX$_{3}$; M = Fe, Ni, Mn, etc.; X = S and Se) in antiferromagnetic (AFM) state \cite{Gong2017,Lee2016,Li2014,Lancon2016,Wildes2015,Wildes2012,Wildes2006,Wildes1998,Wildes2017} have also attracted tremendous interests due to their tunable electronic properties and the potential applications in magnetic materials and spintronic devices \cite{LiX2014,LiX2013}.
These compounds are connected by vdW interactions and bring both magnetism and correlated electron physics to the playground of vdW materials \cite{BREC1986,Chittari2016,Rabu2003,Wang2018}.

The MPX$_{3}$ compounds are typical 2D antiferromagnetic materials, the axial distortion of the MX$_{6}$ octahedron is likely to produce anisotropy that can dominate the spin dimension (Ising, XY or Heisenberg model). For example, FePS$_{3}$ uses the Ising model, NiPS$_{3}$ uses the XY model, and MnPS$_{3}$ uses the Heisenberg model \cite{Lee2016,Wildes2015,Joy1992}. The magnetic exchange of these compounds is mainly super-exchange interaction, and the sign and property of the super-exchange interaction depend on the M-S-M angle of the metal ion and the electron occupancy rate. Also, some MPX$_{3}$ materials exhibit more than one magnetic phase transition with changes in temperature. Therefore, it is of great significance to extend the performance of these MPX$_{3}$ materials to the field of spintronics.

\begin{figure*}[t]
\includegraphics[width=16cm]{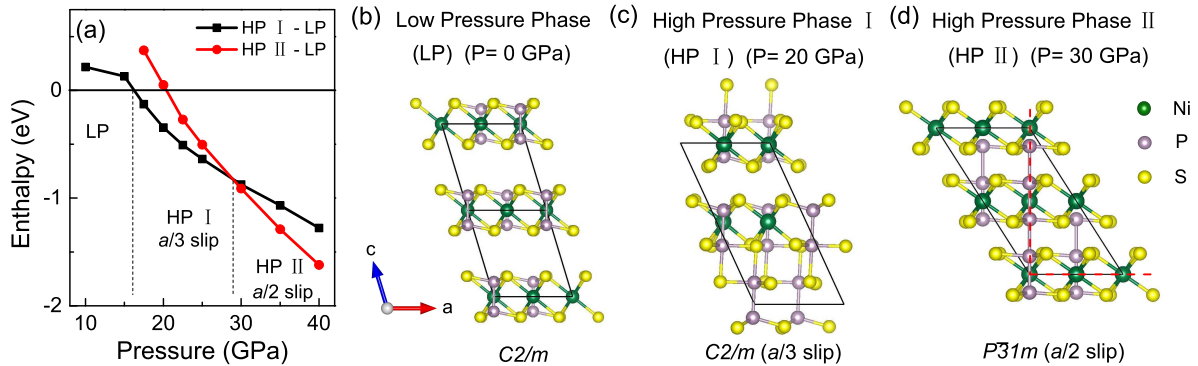}
\caption{(Color online)
Scheme of the NiPS$_{3}$ structure evolution under high pressure. The Ni atoms are shown in green, the P atoms are shown in purple, and the S atoms are shown in yellow. (a) Enthalpy of different phases relative to LP under pressure. (b) LP structure for NiPS$_{3}$ with space group $C2/m$ (view along {\it b}-axis). (c) HP \uppercase\expandafter{\romannumeral1} structure for NiPS$_{3}$ with space group $C2/m$, which slides {\it a}/3 along the {\it a}-axis relative to the LP phase. (d) HP \uppercase\expandafter{\romannumeral2} structure for NiPS$_{3}$ with space group $P\bar{3}1m$, which slides {\it a}/2 along the {\it a}-axis relative to the LP phase. The figures were created using the VESTA software.
}
\end{figure*}

In addition to magnetism, MPX$_{3}$ compounds also have rich electronic properties. Many of these materials are Mott or Charge-Transfer insulators with a wide range of electronic band gaps from 0.25 eV to 3.5 eV \cite{Wang2018,Ouvrard1985,Du2016,Kuo2016,Kurita1989,Whangbo1985,Piacentini1982,Brec1979}, and can be tuned by changing parameters. For example, in the MoS$_{2}$ system, the monolayer MoS$_{2}$ changes from a direct bandgap to an indirect bandgap semiconductor under high pressure, while the bulk material MoS$_{2}$ undergoes a metallic transition under high pressure \cite{Fan2015}. In addition, there are many research groups that have conducted many high-pressure studies on MPX$_{3}$ and reported or predicted the insulator-metal transition \cite{Tsurubayashi2018,Haines2018,Wang2016,WangY2018,Evarestov2020,Kim2019,Coak2019,Coak2020}. 

With the rapid development of 2D materials, high pressure has already become a very important parameter to tune the states of materials and study their fundamental physical properties, such as structural phase transition, electronic behavior and even magnetism. For example, as mentioned above,  at sufficient pressure, the band gap of many materials can close and experience insulator-metal transition. Many novel and rich physical properties can be obtained through pressure regulation. MPX$_{3}$ provide a good platform for us to study how the phases and interactions evolve when the system is tuned from the 2D limit to the 3D state through pressure.

In this paper, we report a systematic study on the structure, electronics and magnetism of NiPS$_{3}$ under pressure.
A two-stage phase transition which undergoes 2D-3D crossover is found in bulk NiPS$_{3}$ by $ab$ $initio$ calculations, high-pressure XRD and Raman measurements. During the whole structure phase transition, the interlayer distance changes more obviously compared with other unit cell parameters. The first structural phase transition occurs at about 15 GPa, and the new structure has the same space group ($C2/m$) as the ambient structure with a {\it a}/3 slip along the {\it a}-axis. Meanwhile, the magnetism collapse and insulator-metal transition in pressurized bulk NiPS$_{3}$ are observed around 20 GPa through high-pressure resistance, infrared, Raman and other measurements. The second structural phase transition occurs at about 27 GPa with a {\it a}/2 slip along the {\it a}-axis, and its space group becomes $P\bar{3}1m$, exhibiting 3D behavior.

\section{RESULTS AND DISCUSSION}
\subsection{Structure phase transition}
At ambient condition, NiPS$_{3}$ has a layered structure in monoclinic ($C2/m$, No. 12) symmetry \cite{Ouvrard1985,Klingen1968}.
In order to explore the structure changes of NiPS$_{3}$ under pressure, we have performed a detailed $ab$ $initio$ simulation using a double cell along the $c$-axis [see Fig. 1(b)].
Several possible structures are considered via a layer by layer slip mechanism along the $a$-axis.
Fig. 1(a) displays the enthalpy change of different phases relative to the original monoclinic phase under pressure.
According to the principle of lowest energy, we confirm three phases up to 40 GPa. We temporarily define the structure at 0 GPa as low pressure phase (LP); the phase at 20 GPa is defined as high pressure phase \uppercase\expandafter{\romannumeral1} (HP-I); the one at 30 GPa is defined as high pressure phase \uppercase\expandafter{\romannumeral2} (HP-II).
The LP phase is existing below 16 GPa, HP-II phase is more faverable above 28 GPa,
and HP-I phase is located between 16$\sim$28 GPa, showing as an intermediate state.
Fig. 1(b) displays the crystal structure of the LP phase of NiPS$_{3}$. Viewed along {\it b}-axis, we can see the well separated sandwich layers associated with weak interlayer vdW interaction. The HP-I structure has a {\it a}/3 slip along {\it a}-axis [Fig. 1(c)]. The slip between the {\it ab} planes with pressure is feasible as the planes are only weakly bound by vdW forces and they can slide easily over one another. LP and HP-I phases are isostructural phase transition, they have the same space group $C2/m$. The HP-II structure has a {\it a}/2 slip along {\it a}-axis [Fig. 1(d)]. This phase transition leads to the closely related $P\bar{3}1m$ (No. 162) trigonal structure which adopts a higher symmetry in paramagnetic state (see Fig. 5).
From LP to HP-II phase of the entire structure evolution process, we can clearly see a 2D-3D crossover, relative to the intermediate state of HP-I phase.
These results suggest a two-stage phase transition process from monoclinic ($C2/m$) to trigonal ($P\bar{3}1m$) structure via a layer by layer slip mechanism under pressure.

\begin{figure*}[t]
\includegraphics[width=16cm]{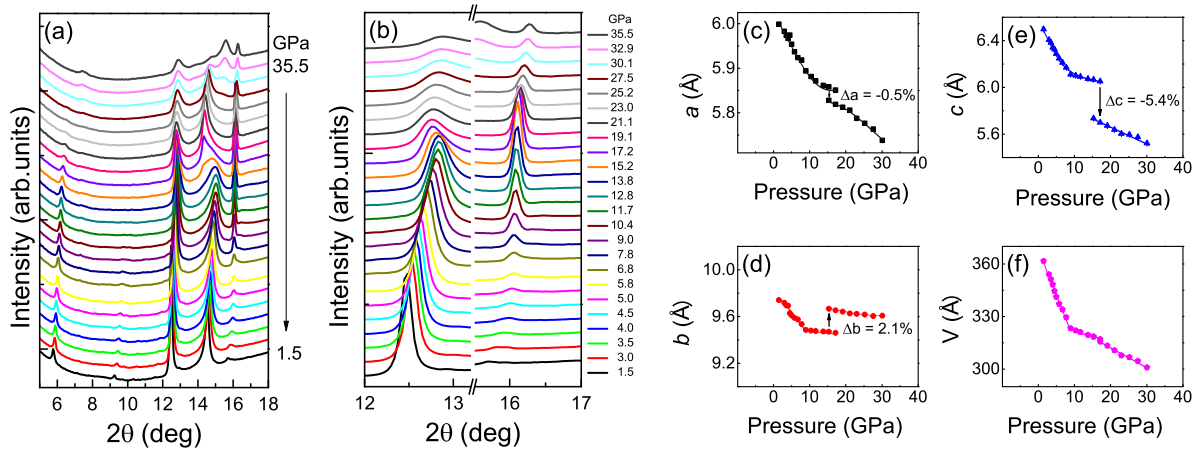}
\caption{(Color online)
In situ x-ray diffraction study on NiPS$_{3}$ powder sample.  (a) The diffraction pattern of NiPS$_{3}$ evolved with pressure , and the x-rays wavelength was $\lambda$ = 0.6199 {\AA}. (b) The detail information of diffraction pattern of NiPS$_{3}$ from 12.0$^{\circ}$ to 16.8$^{\circ}$. (c) Experimental cell parameters {\it a} as a function of applied pressure for the LP and HP phases of NiPS$_{3}$. (d) Experimental cell parameters {\it b} as a function of applied pressure for the LP and HP phases of NiPS$_{3}$. (e) Experimental cell parameters {\it c} as a function of applied pressure for the LP and HP phases of NiPS$_{3}$. (f) Experimental cell volumes as a function of applied pressure for the LP and HP phases of NiPS$_{3}$.
}
\end{figure*}

To verify the calculated structure model, we have conducted a high-pressure XRD measurement of NiPS$_{3}$.
The diffraction patterns evolved with pressure are presented in Fig. 2(a). From the diffraction patterns, two distinct phase transitions can be observed near 15 GPa and 27 GPa, respectively. The first phase transition begins at approximately 15 GPa and completely transitions to another phase around 20 GPa. The second phase transition occurs at approximately 27 GPa. For the first phase transition, there is a new diffraction peak develops while other peak disappears. The evolution of the peak around 14.5$^{\circ}$ begins to occur near 15 GPa. It evolves into a new peak at about 20 GPa and then becomes another phase upon this pressure.
In addition to this change, the other two diffraction peaks $\sim$ 12.5$^{\circ}$ and $\sim$ 16.0$^{\circ}$ [see Fig. 2(b)] also have corresponding responses around 15 GPa.
Below 15 GPa, the peak around 12.5$^{\circ}$ gradually moves to a higher angle as the pressure gradually increases,
while above 15 GPa, it moves to a lower angle first, and then continues to move to a high angle as the pressure increases;
On the other hand, below 15 GPa, the intensity of the peak around 16.0$^{\circ}$ increases with increasing pressure,
while above 15 GPa, the intensity of this diffraction peak weakens with increasing pressure.
Based on the above phenomena, we believe that NiPS$_{3}$ undergoes the first phase transition occurs around 15 GPa.
The phase transition around 15 GPa has also been verified in the Raman scattering measurement (see Fig. S1 in the Supplementary Material \cite{sm00}). For the second phase transition at approximately 27 GPa, there are also new diffraction peaks develop while one peak disappears. The diffraction peak around 6$^{\circ}$ disappears, and two new peaks appear around 6.5$^{\circ}$ and 14.5$^{\circ}$ at approximately 27 GPa.
Our experimental results are well consistent with the calculated results shown in Fig. 1(a).

Fig. 2(c-f) shows the experimental cell parameters as a function of applied pressure for the LP and HP phases of NiPS$_{3}$. From Fig. 2(c-e) we can see that all the lattice parameters $a$, $b$ and $c$ are in a consistent trend when they are below 10 GPa. This interval is the LP phase. In contrast, in the pressure range of 10-15 GPa, the change trend of the unit cell parameters slow down with increasing pressure. This interval should be a transition interval. All unit cell parameters $a$, $b$ and $c$ above 15 GPa change suddenly, and the volume gradually decreases with increasing pressure [Fig. 2(f)]. The lattice parameter {\it a} decreases by 0.5\% [Fig. 2(c)], {\it b} increases by 2.1\% [Fig. 2(d)], and {\it c} decreases by 5.4\% [Fig. 2(e)]. Compared with the unit cell parameters {\it a} and {\it b}, the unit cell parameter {\it c} which includes the vdW gap decreases rather rapidly with the pressure .
So the effect of pressure on the interlayer is greater.
In this interval, a structural phase transition should occur, which is a HP-I phase.
This HP-I and LP phase adopt the same space group of $C2/m$. Thus, monoclinic unit cells with space group $C2/m$ were adopted for cell parameter fitting for NiPS$_{3}$. However, because the XRD peaks above 30 GPa are weaker, fewer and wider, it is difficult to fit. So the fitting data above 30 GPa is not given.
In addition, because 27-30 GPa belongs to the transition interval, so in this interval we still use the HP-I structure for fitting. It can be found from the fitting result that the change of the unit cell parameters are the same as the result given by the simulation calculation. The obvious decline of the interlayer distance also provides more favorable conditions for 2D-3D crossover.

\begin{figure*}
\includegraphics[width=16cm]{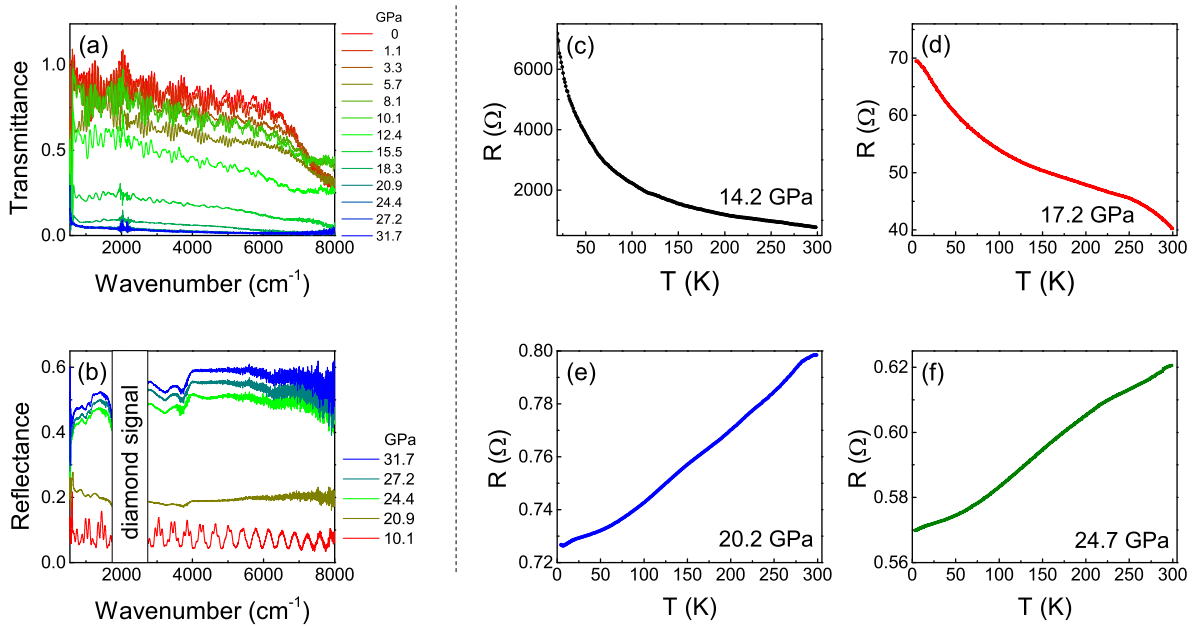}
\caption{(Color online)
Electronic behavior of bulk NiPS$_{3}$ crystal under high pressure. (a-b) The infrared measurement of bulk NiPS$_{3}$ under different pressures: (a) transmittance. (b) reflectance. (c-f) Temperature dependent  resistance of NiPS$_{3}$: 14.2 GPa, 17.2 GPa, 20.2 GPa, and 24.7 GPa, respectively. An insulator-metal transition occurred at about 20 GPa.
}
\end{figure*}

\subsection{Insulator-Metal transition}
The infrared measurement has important guiding significance for us to understand the high-pressure electronic behavior of materials, therefore, we performed infrared measurement on bulk NiPS$_{3}$ under high pressure. The high-pressure infrared measurement includes two parts: transmittance and reflectance. The transmittance of the insulator is close to 1, and the transmittance of metal is close to 0. As can be seen from Fig. 3(a), under ambient pressure, the transmittance of NiPS$_{3}$ is close to 1, which is an insulator. As the pressure gradually increases, the transmittance of NiPS$_{3}$ gradually decreases, indicating that its band gap is gradually decreasing, and the insulator gradually transitions to metal. When the pressure increases to 20.9 GPa, the transmittance of NiPS$_{3}$ is close to 0, and it becomes a metal. As the pressure continues to increase, the transmittance of NiPS$_{3}$ no longer changes. In addition, as the reflectance increases, the metallicity will increase, so we continue to observe the metallization process through reflectance. By observing the change of reflectance under different pressures, we also got the same conclusion. From Fig. 3(b), we can see that the reflectance is small and the interference is strong at 10 GPa. As the pressure gradually increases, the reflectance increases, and the spectrum gradually smoothes. At around 20.9 GPa, the spectrum becomes very smooth , suggesting that there is no interference between top surface and bottom surface of NiPS$_{3}$ crystal, which is a typical character of metallic behavior. The reflectance increases significantly when pressure reaches 24.4 GPa, above which reflectance tends to be stable. The signal around 2000 cm$^{-1}$ is the absorption signal of diamond. In addition, we also clearly observed the metallization process by optic reflection picture of NiPS$_{3}$. For details, see Fig. S2 in Supplementary Material \cite{sm00}.

In order to verify the accuracy of the metallization behavior, we measured the resistance of bulk NiPS$_{3}$ under high pressure. Temperature-dependent resistance under different pressures are plotted in Fig. 3(c-f). A transition from insulator to metal is seen around 20 GPa. From Fig. 3(c-f) we can see that as the pressure gradually increases, the resistance is gradually suppressed, and there is a decrease in magnitude. For details, see Fig. S3 in Supplementary Material \cite{sm00}.
Through Fig. 3(c-f) we can see the metallization process of NiPS$_{3}$. At 14.2 GPa the data appears as an insulator and to be an intermediate state at 17.2 GPa [Fig. 3(c) and (d)]. Continue to increase pressure to about 20.2 GPa [Fig. 3(e)], the sample completely metallized. We continue to increase the pressure until 24.7 GPa [Fig. 3(f)], NiPS$_{3}$ still maintains a good metal behavior. This measurement confirms the result of infrared measurement.

\begin{figure*}[t]
\includegraphics[width=16cm]{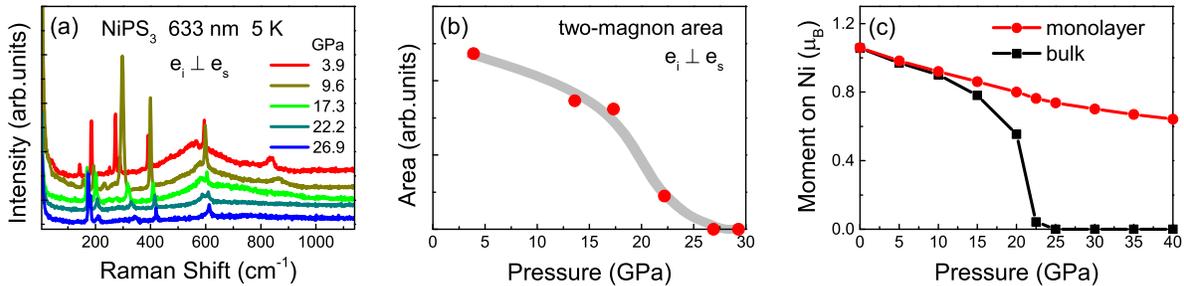}
\caption{(Color online)
Magnetic behavior of NiPS$_{3}$ crystal under high pressure. (a) Raman spectra of bulk NiPS$_{3}$ in cross-polarization configuration under different pressures at 5 K. (b) Two-magnon area of bulk NiPS$_{3}$ in cross-polarization configuration under different pressures, the data points in (b) are extracted from Raman spectra in (a). (c) The moment of NiPS$_{3}$ on Ni under different pressures.
}
\end{figure*}

The above measurement methods have clearly confirmed that the bulk NiPS$_{3}$ will undergo an insulator-metal phase transition under pressure. At present, related articles predict that monolayer MPX$_{3}$ will still maintain insulator or semiconductor behavior under pressure without metallization \cite{Xiang2016}. That is, the bulk NiPS$_{3}$ exhibits 3D characteristics that are different from "pure" 2D (monolayer) NiPS$_{3}$. We speculate that the difference in behavior between bulk NiPS$_{3}$ and monolayer NiPS$_{3}$ may be caused by the interlayer interaction.

\subsection{Magnetic property}
In order to verify the correlation between the magnetic properties and the interlayer interaction of NiPS$_{3}$, we have measured and calculated the magnetic properties of NiPS$_{3}$. By performing temperature-dependent Raman scattering measurements on bulk NiPS$_{3}$ from 5.6 to 300 K at ambient pressure, we discovered the two-magnon signal of NiPS$_{3}$ centered at ~550 cm$^{-1}$. Details can be seen in Fig. S4 in the Supplementary Material \cite{sm00}. Consistent with the results measured in other articles \cite{Rosenblum1999,KimK2019}. In order to investigate the effect of pressure on two-magnon, we conducted the Raman measurements of bulk NiPS$_{3}$ in the cross-polarization configuration under different pressures at the low temperature of 5 K [Fig. 4(a)]. In the cross-polarization configuration, some phonons, especially those superimposed on the two-magnon, are suppressed, so the two-magnon signal in this configuration is clearer. From these data we can find that the two-magnon signal was gradually suppressed and shift towards higher frequencies as the pressure increased. The two-magnon signal became very small around 22 GPa and was totally suppressed around 27 GPa. Such a rule is also satisfied under different polarization configurations and different temperatures. For details, see Fig. S5 and S6 in Supplementary Material \cite{sm00}. In order to quantitatively describe the change of magnetism with pressure, we fitted the data in Fig. 4(a) and gave the two-magnon area of bulk NiPS$_{3}$ under different pressures [Fig. 4(b)]. From Fig. 4(b), we can see that the area of the two-magnon decreases with increasing pressure. When the pressure reaches about 27 GPa, the magnetism completely disappears.

In addition to the experimental results, we also calculated the magnetic moment of NiPS$_{3}$ under pressures with an intralayer AFM structure \cite{Rao1992}. In Fig. 4(c), the calculated moments on Ni sites are plotted as a function of pressure up to 40 GPa for bulk NiPS$_{3}$. We can see that the moments on Ni sites are quickly deceased from 15 GPa and quenched around 22 GPa. Its behavior is consistent with the trend of the area of the two-magnon changing with pressure in Fig. 4(b). For comparison, we have also calculated the magnetic moments for the monolayer NiPS$_{3}$ (just extend the interlayer spacing in a double cell along the {\it c}-axis) at each pressure.
As shows in Fig. 4(c), the moments on Ni sites almost show a linear decreasing behavior and are well kept up to 40 GPa, indicating that the AFM state is expected to be stable in the monolayer limit under high pressure. Similar to the electrical behavior, the magnetic behavior of bulk NiPS$_{3}$ also shows 3D characteristics which different from "pure" 2D (monolayer) NiPS$_{3}$. These results suggest that the magnetic moments are strongly influenced by the interlayer interaction.

To get the best understanding of interlayer interaction, we have calculated the total and projected electronic density of states (DOS) at 20 GPa for both bulk and monolayer NiPS$_{3}$ (see Fig. S7 in Supplementary Material \cite{sm00}). It is shown that the bulk NiPS$_{3}$ at 20 GPa has a metallic behavior, while the monolayer NiPS$_{3}$ has a semiconducting behavior with a band gap of 0.43 eV.
These results suggest the magnetic moments and electronic property (metallic or semiconducting) are strongly correlated with the interlayer distance.

\begin{figure}
\includegraphics[width=9cm]{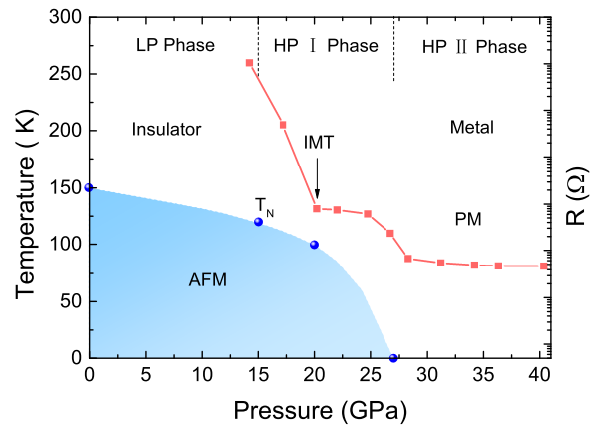}
\caption{(Color online)
Temperature-pressure phase diagram of bulk NiPS$_{3}$. The structural, magnetic  and electronic transition as a function of pressure. The pressures corresponding to the beginning and end of the gradual LP-HP \uppercase\expandafter{\romannumeral1}-HP \uppercase\expandafter{\romannumeral2} structural transition, the insulator-metal transition (IMT) and the AFM-PM transition are marked with dotted lines and arrows. The red points in the figure are the resistance under different pressures at 300 K, and the blue points are the T$_{N}$ under different pressures.
}
\end{figure}

\subsection{Phase diagram}
Fig. 5 shows the temperature-pressure phase diagram of bulk NiPS$_{3}$. Two structural transitions are clearly visible at 15 GPa and 27 GPa, respectively. The structure below 15 GPa is the LP phase. The phase between 15-27 GPa is the HP-I phase, a transition region of 2D-3D. The structure above 27 GPa is the HP-II phase. It can be seen from the phase diagram that the insulator-metal transition happens at the mixed phase (HP-I phase). The red point in the figure is the resistance under different pressures at 300 K. As the pressure increases, the resistance suddenly decreases and metallizes around 20 GPa. The AFM-PM transition ccurs at the junction of HP-I and HP-II phases. As can be seen, T$_{N}$ is suppressed progressively to 0 K at 27 GPa, above which the antiferromagnetic disappears.

\section{CONCLUSION}
In summary, we report the close correlation among the structure, electronic and magnetic properties of NiPS$_{3}$ driven by external hydrostatic pressure. Dimensional crossover appears not just in structural phase transition, but also in electronic behavior and magnetism evolution. Two new phases HP-I and HP-II of NiPS$_{3}$ are proposed by $ab$ $initio$ calculations and confirmed by high-pressure XRD experiment. The change from LP to HP-I structure is accompanied with a {\it a}/3 slip along the {\it a}-axis. The space group is still $C2/m$, which is an isostructural phase transition. Compared with the LP phase, the cell parameters change suddenly in this phase. In particular, the change in the interlayer distance is most obvious. As a result, the magnetism of bulk NiPS$_{3}$ are gradually suppressed, while the magnetism of the monolayer NiPS$_{3}$ remains stable. At the same time, bulk NiPS$_{3}$ completed the transition from insulator to metal in this phase, while monolayer NiPS$_{3}$ is still a semiconductor under high pressure. The structural change from HP-I to HP-II is accompanied with a {\it a}/2 slip along the {\it a}-axis. In the meanwhile, a 2D-3D crossover occurred in structure, forming a new space group $P\bar{3}1m$. NiPS$_{3}$ maintains metallic behavior in this phase. At this time, due to structural changes, the magnetism of NiPS$_{3}$ is completely suppressed, and the magnetic moment on Ni sites becomes 0. Both electrical and magnetic properties show 3D characteristics different from their "pure" 2D behavior under pressure. These results suggest these properties are strongly correlated with the interlayer distance. The properties of this compound under pressure are very rich, which provides a good model for the future research on the physical properties of other layered magnetic materials.

%%\hspace{0.5cm}\\

\section*{ACKNOWLEDGMENTS}

This study was supported by the National Key R\&D Program of China under Grant Nos 2016YFA0401503, 2018YFA0305700, 2017YFA0302904 and 2016YFA0300500, the National Natural Science Foundation of China under Grant No. 11575288, 11974387, U1932215, U1930401 and 11774419, the Strategic Priority Research Program and Key Research Program of Frontier Sciences of the Chinese Academy of Sciences (Grant Nos. XDB33000000, XDB25000000 and Grant No. QYZDBSSW-SLH013) and the Youth Innovation Promotion Association of Chinese Academy of Sciences under Grant No 2016006. ADXRD measurements were performed at 4W2 High Pressure Station, Beijing Synchrotron Radiation Facility (BSRF), which is supported by Chinese Academy of Sciences (Grant KJCX2-SW-N20, KJCX2-SW-N03). The work is partially carried out at High-pressure synergetic measurement station of synergetic extreme condition user facility.

\section*{APPENDIX: MATERIALS AND METHODS}

\subsection{Sample synthesis}

Single crystals of NiPS$_{3}$ were grown by chemical vapor transport method (CVT). The crystal naturally cleaves along the (001) surface, forming NiPS$_{3}$ flakes weakly bonded by van der Waals force. Amount of high-purity elements nickel, phosphorus, and sulfur were mixed in a stoichiometric mole ratio of 1:1:3 (around 1 g in total) and iodine (about 0.123 g) was used as a transport agent. These raw materials were sealed into a quartz ampoule with a length of about 16 cm and an external diameter of 13 mm, and the pressure inside the quartz ampoule was pumped down to 1$\times$10$^{-4}$ Torr. Put the quartz ampoule into a two-zone furnace (680-720 $^{\circ}$C) and heat it for two weeks. Cool the quartz ampoule down to room temperature, and bulk crystals can be obtained on the top of the quartz ampoule.

\subsection{The high-pressure measurements}

The high-pressure four-probe electrical resistance, X-ray powder diffraction (XRD) and Raman measurements were performed in diamond anvil cells (DACs). The DAC made of BeCu alloy with two opposing anvils was used to generated high pressure. Diamond anvils with 300 $\mu${\it m} culets were used for the measurements. In these experiments, a thin single crystal sample of 10*50*100 $\mu${\it m} was loaded into the sample chamber (D = 100 $\mu${\it m} hole) in a rhenium gasket with c-BN insulating layer, and a ruby ball is loaded to serve as internal pressure standard. The ruby fluorescence method was used to determine the pressure. The high-pressure four-probe electrical resistance measurement is carried out at High-pressure synergetic measurement station of synergetic extreme condition user facility \cite{Yu2018}.

The high-pressure XRD measurements were performed at beamline 4W2 at Beijing Synchrotron Radiation Facility. A monochromatic X-ray beam with a wavelength of 0.6199 {\AA} was used for the measurements. Silicone oil was used as pressure-transmitting medium.

The high-pressure and temperature-dependent Raman spectra were collected using a HR800 spectrometer (Jobin Yvon) equipped with a liquid-nitrogen-cooled charge-coupled device (CCD) and volume Bragg gratings, for which micro-Raman backscattering configuration was adopted. A 633 nm laser was used, with a spot size of ~5 $\mu${\it m} focused on the sample surface. The laser power was maintained at approximately 1.4 mW to avoid overheating during measurements. KBr was used as pressure-transmitting medium.

The high-pressure infrared experiments were performed at room temperature on a Bruker VERTEX 70v infrared spectroscopy system with HYPERION 2000 microscope. A thick NiPS$_{3}$ single crystal was used and KBr was used as pressure medium. The spectra were collected in transmission and reflective mode in the range of 600-8000 cm$^{-1}$ with a resolution of 4 cm$^{-1}$ through a $\sim$30$\times$30 $\mu${\it m}$^{2}$ aperture.

\subsection{Computational methods}

The calculations are performed using the Vienna $ab$ $initio$ simulation package (VASP) \cite{Kresse1996} with the projector augmented wave method \cite{Blochl1994} and spin polarized generalized gradient approximation \cite{Perdew1996} for the exchange-correlation energy. The valence states 3d$^{9}$4s$^{1}$ for Ni, 3s$^{2}$3p$^{3}$ for P, and 3s$^{2}$3p$^{4}$ for S are used with the energy cutoff of 550 eV for the plane wave basis set. To simulate the interlayer interaction, a 1$\times$1$\times$2 supercell (in AA stacking including 8 NiPS$_{3}$ formula) is used with a vdW correction \cite{Tkatchenko2009}. An intralayer zigzag antiferromagnetic structure \cite{Rao1992} is set with the interlayer ferromagnetic coupling. The Brillouin zone is sampled with a 7$\times$4$\times$4 Monkhorst-Pack special k-point grid. Throughout our calculations with or without pressure, convergence criteria employed for both the electronic self-consistent relaxation and the ionic relaxation were set to 10$^{-6}$ eV and 0.01 eV/A for energy and force, respectively.

%\section{References}

\end{document}